\documentclass{JHEP3}

\usepackage{graphicx}
\usepackage{amsfonts}
\usepackage{amsmath}
\usepackage{amssymb}
\usepackage{amsmath,amssymb,epsfig}
\numberwithin{equation}{section}
\usepackage{cite}

\newcommand{\be}{\begin{equation}}
\newcommand{\bea}{\begin{eqnarray}}
\newcommand{\eea}{\end{eqnarray}}
\newcommand{\ba}{\begin{array}}
\newcommand{\ea}{\end{array}}
\newcommand{\ee}{\end{equation}}

\title{Gauge extensions of supersymmetric models and hidden valleys}

\author{Mingxing Luo and Sibo Zheng \\
Zhejiang Institute of Modern Physics, Department of Physics,\\
Zhejiang University, Hangzhou 310027, P. R. China.\\
 E-mail: \email{luo@zimp.zju.edu.cn}, \email{sibozheng.zju@gmail.com}}

\abstract{Supersymmetric models with extended group structure beyond the standard model are revisited in the framework of general gauge mediation.
Sum rules for sfermion masses are shown to depend genuinely on the group structure,
which can serve as important probes for specific models.
The left-right model and models with extra $U(1)$ are worked out for illustrations.
If the couplings of extra gauge groups are small,
supersymmetric hidden valleys of the scale $10-100$ GeV
can be naturally constructed in companion of a TeV-scale supersymmetric visible sector.}

\begin{document}

\section{Introduction}

Recently, a general method has been proposed to calculate soft terms in gauge mediation models (GGM) \cite{MSS}.
It turns out that all soft terms in a specific model can be determined by a few parameters,
which encode the information of the hidden sector.
One obtains two sum rules \cite{sum_rules} which make a distinctive feature in $R$-symmetry breaking gauge mediation,
in comparsion with gravity mediation \cite{gravity,gravity2},
if the breaking of supersymmetry is communicated to the visible sector only by standard model gauge interactions.
In principle, the hidden sector can be either weakly or strongly coupled.
And the formalism is valid for both direct and non-direct gauge mediation.

In this paper, we will reconsider the gauge extended supersymmetric model (ESM) using the GGM formalism.
The ESM can easily be constructed in deformed ISS theories \cite{0812.3119}.
In these models, the unbroken global symmetry $G_{0}$ in the hidden sector is larger than $ G_{SM}$.
If there are extra gauge structures beyond the standard model (SM)
and the corresponding symmetry in $G_{0}$ is weakly gauged at the supersymmetric breaking scale $M_{SUSY}$,
extra interactions in addition to those of $ G_{SM}$ yield modifications to the soft terms, among other things.
In general, they modify the sum rules in \cite{sum_rules}.
The extra gauge structure is assumed to be spontaneously brokon at an intermediate scale between $M_{SUSY}$ and $M_{EW}$,
with the corresponding gauge boson of masses in the order of a few TeV.
At the electro-weak scale, only the SM is left in the visible sector.
Such arrangement is similar to the $Z'$-mediated supersymmetric breaking \cite{Langacker,Langacker2} in spirit.

To be concrete, we will consider the abelian case where the gauged part of $G_{0}$ is $G_{SM}\times U(1)'$,
and the non-abelian supersymmertic left-right model
where the gauged part of $G_{0}$ is $SU(3)_{c}\times SU(2)_{L}\times SU(2)_{R} \times U(1)_{B-L}$.
We find that the sum rules in \cite{sum_rules} cannot be retained in either cases.
The resulting modifications are easily obtained and dependent on the couplings of extra gauge sector.
If the extra gauge couplings are comparable with the ones in the SM,
the sum rules are broken significantly.
If these couplings are small enough, these sum rules can survive as approximations.
These analysis can be directly applied to theories with more sophisticated gauge structures,
with similar conclusions.

The small coupling case is then used to construct models with a particular type supersymmetric hidden valley models \cite{hidden1}.
In particular, we will construct a model in which the extra $U(1)'$ communicates between the supersymmetry breaking sector
and a hidden valley sector, which generates supersymmetry breakings in the latter.
Simultaneously, the same $U(1)'$ communicates the hidden valley sector to the visible sector.
If the $U(1)'$ coupling is of the order $10^{-1}-10^{-2}$ at $M_{SUSY}$,
the soft terms in the hidden valley are two or four orders of magnitude smaller than those in the visible sector,
which implies that an $\mathcal{O}(1-10)$ TeV-scale visible sector is accompanied by an $\mathcal{O}(10-100)$ GeV-scale hidden valley.

The rest of the paper is organized as follows.
In section II, we briefly review and comment on the GGM formalism.
In section III, we discuss supersymmetric models with group structure beyond SM.
Particular attention will be payed to the sum rules for sfermion masses.
In section IV, we propose a class of supersymmetric hidden valleys with $U(1)'$.
Finally we conclude in section V.

\section{Review and comments on GGM}
In this section, we briefly review the GGM formalism,
with an emphasis on calculations of soft terms and sum rules for sfermion masses in the MSSM.
The gauge current of the hidden sector is a real linear superfield,
\begin{eqnarray}{\label{current}}
\bar{D}^2 \mathcal{J} = D^2\mathcal{J} = 0
\end{eqnarray}
with $\partial_{\mu}j^{\mu}=0$, as required by the current conservation condition.
The two-point correlator of $\mathcal{J}$ can generally be written as,
\begin{eqnarray}\label{correlator}
<\mathcal{J}(p,\theta,\bar{\theta})\mathcal{J}(p',\theta',\bar{\theta}')>
=2\pi^{4}\delta^{4}(p+p')I(p,p',\theta,\theta')
\end{eqnarray}
\eqref{current} and \eqref{correlator} can be generally solved, by a set of real functions $B_{1/2},C_{a}$ \cite{Distler}.
The Fourier transforms of the correlators in momentum space are,
\begin{eqnarray}{\label{correlators}}
 \langle J(p) J(-p)\rangle &=&\tilde C_0(p^2/M^2;\,M/
 \Lambda)\nonumber\\
 \langle j_{\alpha}(p) \bar j_{\dot\alpha}(-p) \rangle
 &=&-\sigma_{\alpha\dot\alpha}^\mu p_\mu \tilde C_{1/2}
 (p^2/M^2;\,M/\Lambda)\nonumber\\
 \langle j_\mu(p) j_\nu(-p) \rangle &=& -(p^2\eta_{\mu\nu}
 -p_\mu p_\nu)\tilde C_1(p^2/M^2;\,M/\Lambda)\nonumber\\
 \langle j_\alpha(p)j_\beta(-p)\rangle &=&
 \epsilon_{\alpha\beta}M \tilde B_{1/2}(p^2/M^2)
\end{eqnarray}
If SUSY is unbroken, $\tilde C_0=\tilde C_{1/2}=\tilde C_1$, $\tilde B_{1/2}=0$.
If supersymmetry is broken, these relations do not hold in general,
as now $(Q_{\alpha}+Q'_{\dot{\alpha}})I\neq 0$ and $(\bar{Q}_{\alpha}+\bar{Q}'_{\dot{\alpha}})I\neq 0$.

The gauge current superfield acts as a source for visible vector superfield via the coupling,
\begin{eqnarray}{\label{coupling}}
\mathcal{L}_{int} =2g \int d^4\theta \mathcal{J} V + \dots = g( J
D -\lambda j -\bar\lambda\bar j-j^\mu V_\mu) + \cdots,
\end{eqnarray}
Note that in writing \eqref{coupling} the Wes-Zumino gauge has been chosen for the vector superfield.
Integrating out the messenger sector, we obtain the effective Lagrangian for the gauge supermultiplet,
\begin{eqnarray}
 \delta\mathcal{L}_{eff} &=& \frac{1}{2} g^2\tilde C_0(0)D^2 -g^2\tilde
C_{1/2}(0)i\lambda\sigma^\mu\partial_\mu\bar\lambda -{1\over4}
g^2\tilde C_1(0)F_{\mu\nu}F^{\mu\nu}\nonumber\\
 &-&{1\over2}g^2(M\tilde B_{1/2}(0)\lambda\lambda+c.c.)+ \dots
 \end{eqnarray}
This gives contributions to the gaugino and sfermion masses, respectively,
\begin{equation}{\label{mass}}
M_{r}= g_{r}^{2}M\tilde B_{1/2}^{(r)}(0),~~~~~~
\tilde{m}_{f}^{2}=\sum_{r=1}^{r=3}g_{r}^{4}c_{2}(f;r)A_{r}
\end{equation}
where $c_2(f;r)$ is the Casimir of the representation $ f$ under the $r$ gauge group and
\begin{eqnarray}{\label{Ar}}
A_{r}=-\int\frac{d^{4}p}{(2\pi)^{4}}\frac{1}{ p^{2}}\left(3
 \tilde{C}_{1}^{(r)}(p^{2}/M^{2})-4\tilde{C}_{1/2}^{(r)}(p^{2}/M^{2})+\tilde{C}_{0}^{(r)}(p^{2}/M^{2})\right)
\end{eqnarray}
Note that the $\mu$ and $B\mu$ terms are model dependent.
They can not be determined unless more assumptions on the Higgs sector and the hidden sector are made.

Now a few comments are in order for \eqref{mass}.
First, one can choose the superfield formalism at the starting point \eqref{coupling}.
The correlator of vector superfields takes a simple form
$<\mathcal{V}\mathcal{V}>=\delta^{4}(\theta-\theta')/p^{2}$ with the gauge fixing parameter ${\xi}=1$.
The wave function renormalization $\mathcal{Z}_{Q}$ in the Kahler potential $\int d^{4}\theta \tilde{Q}e^{-2\mathcal{V}}Q$
yields exactly the soft sfermion masses in \eqref{mass}.
Furthermore, the tri-linear $A$ terms in superpotential can be obtained by replacing $Q$ with canonically normalized $Q'$,
\begin{eqnarray}{\label{new variable}}
Q'=\left(1-\frac{1}{2}\mathcal{Z}|_{\theta^{2}}\theta^{2}-\frac{1}{2}\mathcal{Z}|_{\theta^{2}}\bar{\theta}^{2}\right)Q,
\end{eqnarray}
Second, vector superfileds in \eqref{coupling} are usually massive after the corresponding gauge symmetries are broken.
If $m_{V}>>M_{SUSY}$, they can be integrated out and will play a minor role in the communication of supersymmetry breaking.
If $m_{V}\sim M_{SUSY}$, the effects of $m_{V}$ need then to be taken into account.
At the leading order, one would have
\begin{eqnarray}{\label{A2}}
A'_{r}=-\int\frac{d^{4}p}{(2\pi)^{4}(p^{2}-m_{V}^{2})}\left(\frac{3p^{2}}{p^{2}-m_{V}^{2}}
 \tilde{C'}_{1}^{(r)}(p^{2}/M^{2})-4\tilde{C'}_{1/2}^{(r)}(p^{2}/M^{2})
 +\tilde{C'}_{0}^{(r)}(p^{2}/M^{2})\right)\nonumber\\
\end{eqnarray}
In this paper, we will assume $m_{V}<< M_{SUSY}$ for simplicity
and \eqref{Ar} will be used.

The positivity of $A_{r}$ has been proved in F-term supersymmetry
breaking with $F\ll M^{2}$ \cite{Intriligator,Distler}, where
$A_{r}$ can be written as a derivative term.
From \eqref{Ar}, one sees that there are three independent functions for all sfermion masses in a generation.
Thus, there are at least two independent sfermion masses relations, or sum rules,
\begin{eqnarray}{\label{sumrule}}
Tr\left(Y\tilde{m}_{f}\right)=0,~~~~Tr\left((B-L)\tilde{m}_{f}\right)=0
\end{eqnarray}
These sum rules are one of the most distinctive features in such a gauge mediation setting,
in comparsion with other mediation mechanisms.

\section{Sum rules in ESM}

In this section, we will discuss supersymmetric models with gauge groups
beyond the SM ones $G_{SM}=SU(3)_{c}\times SU(2)_{L}\times U(1)_{Y}$.
We will start with the simple extension with an extra abelian $U(1)'$,
then move on to non-abelian gauges.
In particular, we will concentrate on the left-right symmetric model,
though our analysis can be easily generalized to any theories with more elaborated groups,
with similar conclusions.

\subsection{Abelian case}

Since most of phenomenological results in this section are independent of the details in the hidden sector,
we will not address the issue of realizations of these gauge structure in this section.
In literature, there have been extensive efforts to construct viable models.
For example, gauge extended models in ISS-like theories have been discussed before \cite{deform1,deform2,deform3}.
Theories with similar gauge structures in the hidden sector can be found in \cite{0812.3119},
where the ISS superpotential is deformed by $W_{def}$.\footnote{
 Strongly coupled ISS-like SQCD theories can be described by weakly coupled magnetic dual theories at low energy scale.
The magnetic theories have superpotentials of the same structure as that of generalized O'Raifeartaigh models.
According to the general proof in generalized O'Raifeartaigh models \cite{Shih},
the $R$-symmetry must be spontaneously broken
when $W_{def}$ comes from a set of singlet fields with $R$-charges of neither zero or two.}
On the other hand, the hidden sector discussed in section 4 will be in another paradigm \cite{Giudice},
instead of direct gauge mediation.
Partly, it is because that there are generally unacceptable light gauginos or LHC unaccessible
heavy sfermions in direct gauge mediation, as discussed in \cite{Zohar}.

Here, we assume that there is an extra abelian $U(1)'$ in both the hidden and the visible sectors.
The soft terms can be obtained by calculations similar to the ones in the previous section.
The $U(1)'$ introduces extra $C'_{a}$'s (thus $A'$) and $\tilde{B}'_{1/2}$, which modify the sfermion and gaugino masses
\begin{eqnarray}{\label{modifications1}}
\delta\tilde{m}^{2}_{f_{i}}=\frac{3}{5}g'^{4}q^{2}_{i}A',~~~~~~~
\delta\tilde{M}_{\lambda_{i}}=g'^{2}M\tilde{B}'_{1/2}.
\end{eqnarray}
where $q_{i}$ are the $U(1)'$ charges of fermions and $g'$ is the gauge coupling.
Putting everything together, the soft masses are,
\begin{eqnarray}{\label{soft}}
\left(%
\begin{array}{c}
  m^{2}_{Q} \\
  m^{2}_{U} \\
  m^{2}_{D} \\
  m^{2}_{L} \\
  m^{2}_{E} \\
\end{array}%
\right)=\frac{1}{60}
\left(%
\begin{array}{ccccc}
  80  & 45 & 1  & 36q^{2}_{Q} \\
  80 & 0 &  16 & 36q^{2}_{U} \\
  80 & 0 & 4  & 36q^{2}_{D} \\
  0 & 45 & 9  & 36q^{2}_{L} \\
  0 & 0 & 36  & 36q^{2}_{E} \\
\end{array}%
\right)\left(%
\begin{array}{c}
  g_{3}^{4}A_{3} \\
  g_{2}^{4}A_{2} \\
  g_{Y}^{4}A_{Y} \\
  g^{'4}A' \\
\end{array}%
\right)
\end{eqnarray}
So the sum rules in the previous subsection is not valid in general. Instead, one has
\begin{eqnarray}{\label{sumrule1}}
Tr\left(Y\tilde{m}_{f}\right)&=&
\frac{3}{5}g'^{4}(q^{2}_{Q}-2q^{2}_{U}+q^{2}_{D}-q^{2}_{L}+q^{2}_{E})A'\\
Tr\left((B-L)\tilde{m}_{f}\right)&=&
\frac{3}{5}g'^{4}(2q^{2}_{Q}-2q^{2}_{U}-q^{2}_{D}-2q^{2}_{L}+q^{2}_{E})A'
\end{eqnarray}
Without the $U(1)'$ interaction, one gets back the original sum rules \eqref{sumrule}.
One the other hand, there are five soft masses and four independent $A$'s in \eqref{soft},
from which one can deduce one sum rule for the sfermion masses,
\begin{eqnarray}{\label{rule}}
0&=&\left(q^{2}_{U}-q^{2}_{D}-\frac{1}{3}q^{2}_{E}\right)m^{2}_{Q}
+\left(-q^{2}_{Q}+q^{2}_{D}+q^{2}_{L}-\frac{1}{3}q^{2}_{E}\right)m^{2}_{U}\nonumber\\
&+&\left(q^{2}_{Q}-q^{2}_{U}-q^{2}_{L}+\frac{2}{3}q^{2}_{E}\right)m^{2}_{D}
-\left(q^{2}_{U}-q^{2}_{D}-\frac{1}{3}q^{2}_{E}\right)m^{2}_{L}\nonumber\\
&+&\frac{1}{3}\left(q^{2}_{Q}+q^{2}_{U}-2q^{2}_{D}-q^{2}_{L}\right)m^{2}_{D}
\end{eqnarray}

Now we have a few comments on these results:
\begin{itemize}
 \item The sum rule $Tr(Ym^{2})=0$ holds provided that,
\begin{eqnarray}{\label{condition1}}
q_{E}=q_{D}=0,~~~q^{2}_{Q}=2q^{3}_{U}+q_{L}^{2}
\end{eqnarray} while the other sum rule $Tr((B-L)m^{2})=0$ holds provided that,
    \begin{eqnarray}{\label{condition2}}
     q_{Q}=q_{U}=0,~~~q^{2}_{D}=q^{2}_{E}-2q_{L}^{2}
     \end{eqnarray}
 From \eqref{condition1} and \eqref{condition2},
one can see the original sum rules \eqref{sumrule} cannot be retained at the same time,
except for all $U(1)'$ charges being set to zero.
\item If the visible and the hidden sectors are assumed to be anomaly free separately,
neither sum rules in \eqref{sumrule} can be retained.
 \item   If the coupling $g'$ is substantially smaller than the SM couplings in magnitude,
\eqref{sumrule} hold approximately.
\end{itemize}

The spontaneous breaking of $U(1)'$ can be similar to the usual $U(1)$'s without supersymmetry.
One can introduce standard model singlets $S$ to trigger the breaking
and extra exotic singlets to cancel the anomalies \cite{Langacker}.
In particular, $S$ can obtain vacuum expectation value by radiative corrections,
provided that Yukawa couplings between $S$ and exotic singlets is large enough.

\subsection{Non-abelian case}

We now move on to the discussion of extra non-abelian gauge groups.
For concreteness, we will consider the left-right model with
$G_{0}=SU(3)_{c}\times SU(2)_{L}\times SU(2)_{R}\times U(1)_{B-L}$ \cite{LR},
which breaks into $G_{SM}$ via $SU(2)_{R}\times U(1)_{B-L} \rightarrow U(1)_{Y}$
at some scale $M_R$.
For simplicity, we will assume that $M_R<< M_{SUSY}$,
though our general results do not depend on this assumption.
The analysis can be easily generalized to gauge groups of higher ranks.

The $U(1)_{B-L}$ charges can be easily read from their $U(1)_Y$ charges.
Explicitly, coupling $g_Y$ and charges $q_Y$ are determined by $g_{R},g_{B-L}$ via
\begin{eqnarray}{\label{Y}}
g_{Y}=\frac{g_{R}g_{B-L}}{g_{R}^{2}+g_{B-L}^{2}},~~~~~~
q_{Y,i}=T^{3}_{R,i}+\tilde{q}_{i}.
\end{eqnarray}
It is straightforward to get the masses for soft sfermions,
\begin{eqnarray}{\label{soft2}}
\left(%
\begin{array}{c}
  m^{2}_{Q} \\
  m^{2}_{P} \\
  m^{2}_{L} \\
  m^{2}_{E} \\
\end{array}%
\right)=\frac{1}{60}
\left(%
\begin{array}{ccccc}
  80  & 45 &  0 & 1 \\
  80 & 0 &  45  & 1 \\
  0 & 45 &0  & 9\\
  0 & 0 & 0  & 36 \\
\end{array}%
\right)\left(%
\begin{array}{c}
  g_{3}^{4}A_{3} \\
  g_{L}^{4}A_{L} \\
  g_{R}^{4}A_{R} \\
  g_{B-L}^4A_{B-L} \\
\end{array}%
\right)
\end{eqnarray}
where $P=(U,D)$ carries quantum numbers of $(\mathbf{\bar{3}},\mathbf{1},\mathbf{2},\frac{1}{6})$.

Since fermions in the visible sector fits into spinor representations of $SO(10) \supset G_0$,
it is anomaly free.
So the hidden sector must be anomaly free also.
Generally, there can be chiral matters $S_{i}$ with quantum numbers $(\mathbf{1},\mathbf{1},\mathbf{2},q_{S_{i}})$ $(i\geq 1)$
and $M_{j}$ with quantum numbers $(\mathbf{1},\mathbf{2},\mathbf{1},q_{M_{j}})$ $(j\geq 0)$ in the hidden sector.
The quantum numbers $q$ are constrained by the anomaly free conditions.
Specifically,
\begin{eqnarray}{\label{anomaly4}}
SU(2)_{R}-SU(2)_{R}-U(1)_{B-L}&:&
\sum_{(doublet,S)}\tilde{q}_{i}=0\nonumber\\
SU(2)_{L}-SU(2)_{L}-U(1)_{B-L}&:&\sum_{(doublet,M)}\tilde{q}_{i}=0\nonumber\\
U(1)_{B-L}-U(1)_{B-L}-U(1)_{B-L}&:&\sum_{i=(Q,S,M)}\tilde{q}^{3}_{i}=0\nonumber\\
Graviton-Graviton-U(1)_{B-L}&:&\sum_{i=(Q,S,M)}\tilde{q}_{i}=0
\end{eqnarray}
Other anomaly free conditions are automatically satisfied by the charge assignments in \eqref{Y}.
We note that
\begin{itemize}
    \item The sum rules \eqref{sumrule} are both broken.
    Actually, they are modified to be
    \begin{eqnarray}{\label{rule1}}
    Tr(Ym^{2}_{\tilde{f}})=\frac{3}{4}m^{2}_{E},~~~~~
    Tr((B-L)m^{2}_{\tilde{f}})=\frac{1}{2}m^{2}_{E}
    \end{eqnarray}
These two equations are independent of specific contents of the hidden sector.
Thus, they can serve as important probes of left-right supersymmetric models.
     \item If $SU(2)_{R}\times U(1)_{B-L}\rightarrow U(1)_{Y}$,
     with masses of gauge bosons $(A_{+},~ A_{-},~ A_{0})$ near $M_{SUSY}$,
     the $A_{r}$'s in \eqref{Ar} need to be replaced by those in \eqref{A2}.
     There are then six free parameters $(A_{3},A_{2},A_{Y},A_{+}, A_{-},A_{0})$
     and five sfermion masses.
     This implies \eqref{rule1} is modified again in this case.
     \end{itemize}
The constraints \eqref{anomaly4} can be satisfied by proper assignments of charges $q_{S_{i}}$ and $q_{M_{j}}$.
At least one $S$ is needed to break $G_0$ into $G_{SM}$.

Other extensions of group structure beyond SM induce corresponding sum rules,
some of which can be independent of details of the hidden sector,
which serve as generic probes of such theories.

\section{Supersymmetric hidden valleys}
Usually the hidden sector is assumed to be very heavy.
Actually, a light hidden sector cannot be ruled out
if its communication with the visible sector is sufficiently suppressed.
Scenarios of light hidden sectors with small coupling with the visible sector
has been recently advocated and dubbed as hidden valleys \cite{hidden1}.

In $U(1)$ theories, one always has $\beta_{g'}>0$ and the corresponding couplings decrease
with the decrease of energy.
It is thus possible that the effects from $U(1)'$s are tiny at the electro-weak scale
due to renormalization group flows.
In addition, the couplings between the visible sector and the $U(1)'$s are suppressed further by the massive gauge
boson $m_{Z'}$'s.
So the existence of extra $U(1)'$s cannot be ruled out by present experiments.
Naturally, extra $U(1)'$s has been proposed to communicate the hidden valley sector to the visible sector \cite{hidden1}.

Here we will construct a model in which the extra $U(1)'$ communicates between the supersymmetry breaking sector
and a hidden valley sector, which generates supersymmetry breakings in the latter.
Simultaneously, the same $U(1)'$ communicates the hidden sector to the visible sector.
We will see that if the $U(1)'$ coupling is of the order $10^{-1}-10^{-2}$ at $M_{SUSY}$,
the soft terms in the hidden valley are two or four orders of magnitude smaller than those in the visible sector.
That is to say, an $\mathcal{O}(1-10)$ TeV-scale visible sector is accompanied by an $\mathcal{O}(10-100)$ GeV-scale hidden valley.

To be concrete, we will consider a class of models with the following symmetries and particle contents,
\begin{eqnarray}
\begin{tabular}{|c|c|c|c|c|}
  \hline
      & $SU(n_{v})$ & $G_{SM} $ & $U(1)'$ & $U(1)_{R}$ \\
  \hline
  $X$ & $\mathbf{1}$ & $\mathbf{1}$ & $0$ & $q_X$  \\
  $q_{j}$ & $\Box$ & $\mathbf{1}$ & $q'_{j}$ & 0   \\
  $\Phi_{i}$ & $\mathbf{1}$ & $\Box$ & $q'_{\Phi_{i}}$ & $0$ \\
  $T_{\pm}$ & $\mathbf{1}$ & $\mathbf{1}$ & $q'_{+}+q'_{-}=0$ & $q_{\pm}$  \\
  \hline
\end{tabular}
\end{eqnarray}
Specifically,
\begin{itemize}
    \item The theory is composed of three parts.
    The hidden sector is composed of a spurion $X$ referred to be SUSY-breaking sector and an $SU(n_{v})$ gauge theory with $v$-quarks
    in its bi-fundamental representations.
    The $v$-sector is referred to as hidden valley.
    The messenger sector contains the $\Phi_{i}$'s, which are neutral under $SU(n_{v})$ but charged under $G_{SM}\times U(1)'$.
    The visible sector contains gauged $U(1)'$ extension of group structure beyond $G_{SM}$ below $\sqrt{F_X}$.
    \item The gauge symmetry is $SU(n_{v})\times G_{SM} \times U(1)'$.
    Shown in Table 1 are also the quantum numbers and representations of chiral matters.
\end{itemize}
If the SUSY-breaking sector is realized in the scheme of direct gauge mediation,
there will be unacceptable light gauginos or LHC unaccessible heavy sfermions in general \cite{Zohar}.
Thus, we turn to the old paradigm \cite{Giudice} to realize supersymmetry and R-symmetry breaking.
In such a scheme, it is not necessary to construct the hidden sector explicitly.
One simply assumes that a singlet spurion $X$ is responsible for supersymmetry breaking
and $X$ almost determines all the phenomenological features.
 For explicit SUSY-breaking sectors that induce such a spurion X,
see \cite{Giudice} and reference therein.

The Lagrangian for the model in the table reads,
\begin{eqnarray}{\label{action}}
\mathcal{L}=\int d^{2}\theta W+\int d^{4}\theta K+
\int d^{2}\theta \left(\mathcal{W}_{MSSM}^{2}+\mathcal{W}^{'2}+\mathcal{W}_{h}^{2}\right)
\end{eqnarray}
where
\begin{eqnarray}{\label{action1}}
W&=&\lambda_{ij}X\bar{\Phi}_{i}\Phi_{j},~~~~~~~~~~~~X=M+F_{X}\theta^{2}\\
 K&=&\left(\Phi^{\dagger}_{i} e^{-2V_{MSSM}-2V'}\Phi_{i}+Q_{m}^{\dagger} e^{-2V_{MSSM}-2V'}Q_{m}
+q_{j}^{\dagger}\ e^{-2V_{h}-2V'}q_{j}+T_{\pm}^{\dagger}\ e^{-2V'}T_{\pm}\right)\nonumber\\
\end{eqnarray}
where $Q_m$ denote the chiral matter superfields in SSM sector,
$V_{h}$ and $\mathcal{W}_{h}$ denote the vector and spinor superfield of hidden valley gauge theory respectively.
$T_{\pm}$ are responsible for triggering the spontaneously breaking of $U(1)'$.
\begin{figure}
\centering
\begin{minipage}[b]{0.5\textwidth}
\centering
\includegraphics[width=2.5in]{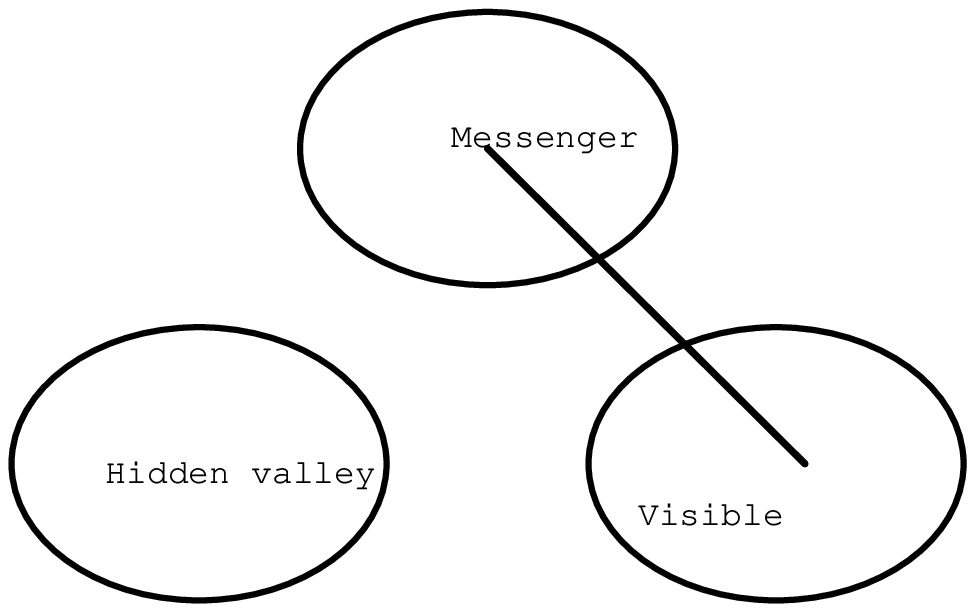}
\end{minipage}%
\begin{minipage}[b]{0.5\textwidth}
\centering
\includegraphics[width=2.5in]{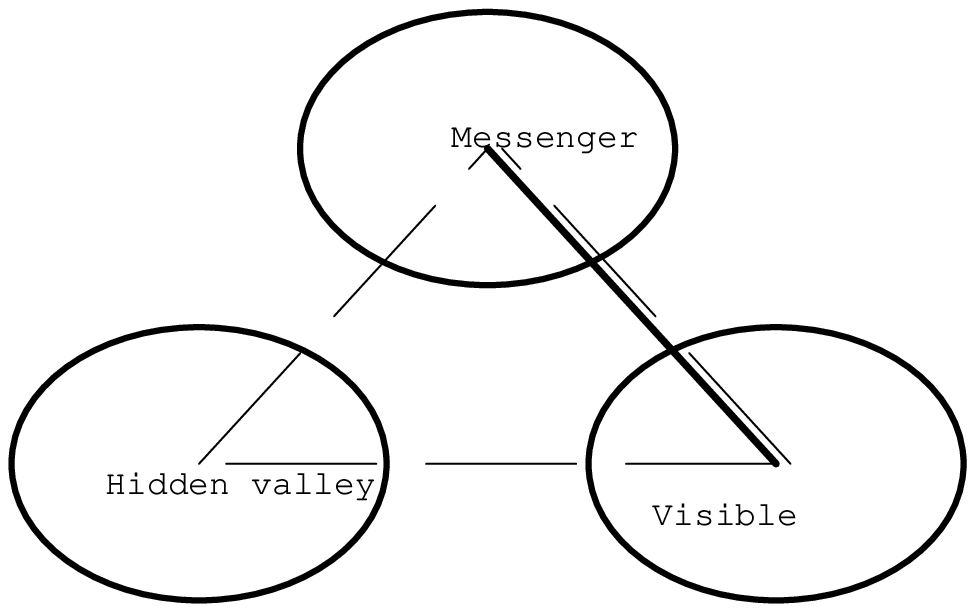}
\end{minipage}
\caption{Gauge mediation with (right) and without (left) extra $U(1)'$.
The black line indicates gauge mediation due to $G_{SM}$ while the dashed ones due to $U(1)'$.}
\end{figure}

One can see any two of the three sectors can communicate through the gauged $U(1)'$ group theory,
as shown in Figure.1.
Both the messenger sector and the visible sector contain $G_{SM}$ gauge interactions,
which dominate the communication between them.
But the hidden valley communicates with others only via the $U(1)'$.

It is straightforward to work out the soft masses in both the visible sector and the hidden valley
($\lambda_{ij}=\delta_{ij}=1$).
Generically,
\begin{eqnarray}{\label{mass1}}
\tilde{m}^{2}_{Q_{i}}&=& \sum_{a=1}^3
\frac{g_{a}^{4} N_{mess}}{(16\pi^{2})^{2}}C_{2}(f_{i},a)\left(\frac{F_{X}}{M}\right)^{2}+\mathcal{O}(g'^{4}) \nonumber\\
M_{\lambda_{a}}&=&\frac{g_{a}^{2}N_{mess}}{(16\pi^{2})}\left(\frac{F_{X}}{M}\right)+\mathcal{O}(g'^{2})
\end{eqnarray}
in the visible sector and
\begin{eqnarray}{\label{mass2}}
\tilde{m}^{2}_{q_{i}}=
\frac{3g'^{4}N_{mess}q'^{2}_{i}}{5(16\pi^{2})^{2}}\left(\frac{F_{X}}{M}\right)^{2},~~~~~
\tilde{m}^{2}_{T_{\pm}}=\frac{3g'^{4}N_{mess}q'^{2}_{\pm}}{5(16\pi^{2})^{2}}\left(\frac{F_{X}}{M}\right)^{2}
\end{eqnarray}
in the hidden valley and $T_{\pm}$ chiral superfields respectively, $N_{mess}$ is the number of messengers.
Finally, the gaugino mass of $U(1)'$ vector superfield reads,
\begin{eqnarray}{\label{mass3}}
M_{\lambda_{\bar{V}}}=\frac{g'^{2}N_{mess}}{(16\pi^{2})}\left(\frac{F_{X}}{M}\right)
\end{eqnarray}

Below the scale where $U(1)'$ is spontaneously broken at scale $\Lambda$ with mass $M_{V'}=4g'^{2}\Lambda^{2}$,
the $V'$ vector superfield can be integrated out,
leaving following couplings in the effective theory at leading order,
\begin{eqnarray}{\label{eff}}
-\frac{1}{4\Lambda^{2}}\int d^{4}\theta \left((1+\tilde{m}^{2}_{T}\theta^{4})
(\sum_{m}q'_{m}Q^{\dagger}_{m}e^{V_{MSSM}}Q_{m}
+\sum_{j}q'_{j}q^{\dagger}_{j}e^{V_{h}}q_{j})^{2}\right)
\end{eqnarray}
where $q'_{m}$ are the $U(1)'$ charges of chiral matters in MSSM.
\eqref{eff} also induces mixing couplings between operators in MSSM and hidden valley,
which are suppressed by the $U(1)'$ gauge bosons mass.

The tree level Higgs masses $m_{H_{u}}$ and $m_{H_{d}}$ in the visible sector are similar to
\eqref{mass1},
\begin{eqnarray}
\tilde{m}^{2}_{H_{u,d}}= \sum_{a=1}^2 \frac{g_{a}^{4}
N_{mess}}{(16\pi^{2})^{2}}C_{2}(H_{u,d},a)\left(\frac{F_{X}}{M}\right)^{2}\sim
\tilde{m}^{2}_{Q}
\end{eqnarray}
As is well-known, the tree-level lightest Higgs mass $m_h$ is always lighter than $m_Z$,
no matter explicit values of  $m_{H_{u}}$ and $m_{H_{d}}$.
This contradicts experimental observations but
$m_h$ can be lifted over $m_Z$ by taking loop corrections into account.
On the other hand, $m_h$ may be further lifted by including higher dimensional couplings in \eqref{eff}.
Explicitly, the correction to potential in visible sector reads,
\begin{eqnarray}{\label{V}}
\delta V=q'_{H_{u}}\tilde{v}H^{\dagger}_{u}H_{u}+q'_{H_{d}}\tilde{v}H^{\dagger}_{d}H_{d}
+\epsilon_{1}(H_{u}^{\dag}H_{u})^{2}+\epsilon_{2}(H_{d}^{\dag}H_{d})^{2}+
\epsilon_{3}(H^{\dag}_{d}H_{u})^{2}
\end{eqnarray}
where
\begin{eqnarray}
\tilde{v}&=&\frac{\tilde{m}^{2}_{T}}{4\Lambda^{2}}\sum_{j}q'_{\tilde{H}_{j}} |<\tilde{H}_{j}>|^2,\nonumber\\
\epsilon_{1}&=&q'^{2}_{H_{u}}\frac{\tilde{m}^{2}_{T}}{4\Lambda^{2}},\nonumber\\
\epsilon_{2}&=&q'^{2}_{H_{d}}\frac{\tilde{m}^{2}_{T}}{4\Lambda^{2}}\nonumber\\
\epsilon_{3}&=&(q'^{2}_{H_{u}}+q'_{H_{u}}q'_{H_{d}}+q'^{2}_{H_{d}})\frac{\mu^{2}}{4\Lambda^{2}}
+q'_{H_{u}}q'_{H_{d}}\frac{\tilde{m}^{2}_{T}}{4\Lambda^{2}}
\end{eqnarray}
Here $<\tilde{H}_{j}>$ are VeVs of scalars in hidden valley.
Linear approximations $F_{H_{u}}\simeq-\mu H_{d}^{\dag}$ and $F_{H_{d}}\simeq-\mu H_{u}^{\dag}$ have been used in above calculations.
For typical parameters $\Lambda\sim 10^{3}$GeV, $\tilde{m}_{T}\sim 10-100$GeV,
 $<\tilde{H}_{j}>\sim 100$GeV and $\mu\sim 200$GeV,
 the corrections to lightest higgs bosons are dominated by $\epsilon_{3}$,
\begin{eqnarray}
\delta_{\epsilon_{3}}m^{2}_{h}\simeq\epsilon_{3}v^{2}\sim \mathcal{O}(10~GeV)^{2}
\end{eqnarray}
This correction is independent of $\tan\beta$.
Thus it contributes significantly at the large $\tan\beta$ limit, as other contributions are usually proportional to $1/\tan\beta$ \cite{Higgs}.

\begin{figure}
\begin{center}
\includegraphics[scale=0.5]{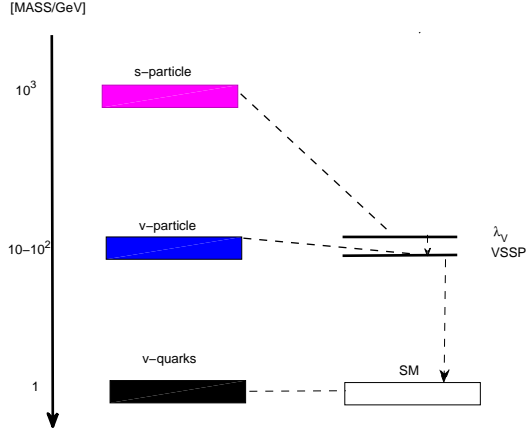}
\caption{Spectra and decay chains of the supersymmetric hidden valley with $F_{X}/M\sim 10^{5}$ GeV.
 The dashed lines refer to particles that decay into
jets/leptons. $\lambda_{\bar{V}}$ and VSSP represent the next-lightest
supersymmetric particles in the visible and hidden sectors respectively.}
\end{center}
\end{figure}

From \eqref{mass1} and \eqref{mass2},
One sees that the soft masses in supersymmetric hidden valley
are two or four order of magnitude smaller than those in visible sector.
Typically, they are in the order of $10-10^{2}$ GeV,
while soft masses of the visible sector and $m_{Z'}$ are usually around TeV.
For such mass parameters, decay chains can be expected between the visible and hidden valley sectors.
In most decaying processes, jets/lepton pairs will be generated, as shown in Figure.2.
Phenomenologically, the generations of $v$-quarks,
the decay widths and their signals at colliders follow the general pattern discussed in \cite{hidden2}.

Finally, we outline the phenomenological features in the visible sector:
\begin{itemize}
 \item As worked out in section 3.1, the sum rules in the visible sector are expected to hold approximately.
 Notice that (most of) results in section 3 are independent of the SUSY-breaking sector.
 \item The gaugino of $U(1)'$ vector superfield is the next-lightest supersymmetric particles
 (NSSP) in the visible sector if $\mid q_{+}\mid$ is larger than $1/\sqrt{0.6N_{mess}}$.
 Otherwise, $T$-scalars are NSSP.
 When sfermion and SM gaugino masses taken to be LHC accessible $\mathcal{O}(1)$ TeV,
 NSSP is around $10-100$ GeV.
 \item At the large $\tan\beta$ limit,
 higher dimensional couplings arising from \eqref{eff} in MSSM are the main sources to correct the Higgs spectra,
 which can substantially uplift the lightest Higgs masses
 across the lower bound at LEPII in the typical parameter space.
\end{itemize}
It would be interesting to construct a single hidden sector,
which spontaneously breaks supersymmetry and $R$-symmetry,
but has desired unbroken gauge symmetry and a hidden valley sector.
One possible realization could be an ISS-like theory with partially unbroken gauge symmetry \cite{deform2}.

\section{Conclusions}

In this paper, we have analyzed supersymmetric models with extended group structure beyond the standard model
in the framework of general gauge mediation.
We have concentrated on the sum rules for sfermion masses,
and they are shown to depend genuinely on the group structure,
which can serve as important probes of the specific model.
In particular, they are rather different from those in models with SM gauge group \eqref{sumrule}.
For definiteness, the left-right model and models with extra $U(1)$ has been worked out in details.
When the couplings of extra gauge groups are smaller than those in the SM,
the sum rules in \eqref{sumrule} hold approximately.

We have constructed a model in which the extra $U(1)'$ communicates between the supersymmetry breaking sector
and a hidden valley sector, which generates supersymmetry breakings in the latter.
Simultaneously, the same $U(1)'$ communicates the hidden sector to the visible sector.
If the $U(1)'$ coupling is of the order $10^{-1}-10^{-2}$ at $M_{SUSY}$,
soft terms in the hidden valley are a few orders smaller than those in the visible sector,
which imply an $\mathcal{O}(1-10)$ TeV-scale visible sector is accompanied by an $\mathcal{O}(1-100)$ GeV-scale hidden valley.
Also, extra higher dimensional couplings help to uplift the mass of the lightest Higgs particle.
The model conforms to the stringent constraints from LEP and other precision experiments,
as the communication between the visible and hidden valley sectors is suppressed
by the massive gauge bosons $m_{Z'}$,
in addition to the smallness of the gauge coupling.

\section*{Acknowledgement}
This work is supported in part by the National Science Foundation
of China (10425525) and (10875103).

\end{document}